\documentclass[graybox]{svmult}
\usepackage{amsfonts}
\usepackage{amsmath}
\usepackage{mathptmx}
\usepackage{helvet}
\usepackage{courier}
\usepackage{type1cm}
\usepackage{makeidx}
\usepackage{graphicx}
\usepackage{multicol}
\usepackage[bottom]{footmisc}

\makeindex

\begin{document}

\title*{Self Sustained Traversable Wormholes and Topology Change Induced by Gravity's Rainbow}
\author{Remo Garattini}

\titlerunning{Traversable Wormholes and Topology Change in Gravity's Rainbow}

\institute{Remo Garattini \at Universit\`{a} degli Studi di Bergamo, Dipartimento di Ingegneria,\\
Viale Marconi,5 24044 Dalmine (Bergamo) ITALY\\
I.N.F.N. - sezione di Milano, Milan, Italy, \email{remo.garattini@unibg.it}}
\maketitle


%
%

\abstract*{We consider the effects of Gravity's Rainbow on the
self-sustained equation which is responsible to find new traversable
wormholes configurations which are sustained by their own gravitational
quantum fluctuations. The same self-sustained equation is also used to
discover if topology change is possible. In this contribution, we will show
that in both uses, the self-sustained equation will produce a Wheeler
wormhole, namely a wormhole of Planckian size. This means that, from the
point of view of traversability, the wormhole will be traversable in
principle, but not in practice. From the topology change point of view, the
background metric will be fixed to be Minkowskian in the equation governing
the quantum fluctuations, which behaves essentially as a backreaction
equation, and the quantum fluctuations are let to evolve. Analyzing this
procedure, we will show that the self-sustained equation, endowed with a
Gravity's Rainbow distortion, will be responsible of a topology change with
the appearance of a Planckian wormhole.}

\abstract{We consider the effects of Gravity's Rainbow on the
self-sustained equation which is responsible to find new traversable
wormholes configurations which are sustained by their own gravitational
quantum fluctuations. The same self-sustained equation is also used to
discover if topology change is possible. In this contribution, we will show
that in both uses, the self-sustained equation will produce a Wheeler
wormhole, namely a wormhole of Planckian size. This means that, from the
point of view of traversability, the wormhole will be traversable in
principle, but not in practice. From the topology change point of view, the
background metric will be fixed to be Minkowskian in the equation governing
the quantum fluctuations, which behaves essentially as a backreaction
equation, and the quantum fluctuations are let to evolve. Analyzing this
procedure, we will show that the self-sustained equation, endowed with a
Gravity's Rainbow distortion, will be responsible of a topology change with
the appearance of a Planckian wormhole.}

\section{Introduction}

\label{sec:1}A wormhole is often termed Einstein-Rosen bridge because a
\textquotedblleft \textit{bridge}\textquotedblright\ connecting two
\textquotedblleft sheets\textquotedblright\ was the result obtained by A.
Einstein and N. Rosen in attempting to build a geometrical model of a
physical elementary "particle" that was everywhere finite and singularity
free\cite{ER}. It was J.A. Wheeler who introduced the term wormhole\cite{JAW}%
, although his wormholes were at the quantum scale. We have to wait for M.
S. Morris and K. S. Thorne\cite{MT} to see the subject of wormholes
seriously considered by the scientific community. In practice a traversable
wormhole is a solution of the Einstein's Field equations, represented by two
asymptotically flat regions joined by a bridge or, in other word, it is a
short-cut in space and time. To exist, traversable wormholes must violate
the null energy conditions, which means that the matter threading the
wormhole's throat has to be \textquotedblleft \textit{exotic}%
\textquotedblright . Classical matter satisfies the usual energy conditions.
Therefore, it is likely that wormholes must belong to the realm of
semiclassical or perhaps a possible quantum theory of the gravitational
field. Since a complete theory of quantum gravity has yet to come, it is
important to approach this problem semiclassically. On this ground, the
Casimir energy on a fixed background. has the correct properties to
substitute the exotic matter: indeed, it is known that, for different
physical systems, Casimir energy is negative. Usually one considers some
matter or gauge fields which contribute to the Casimir energy necessary to
the traversability of the wormholes, nevertheless nothing forbids to use the
Casimir energy of the graviton on a background of a traversable wormhole. In
this way, one can think that the quantum fluctuations of the gravitational
field of a traversable wormhole are the same ones which are responsible to
sustain traversability. Nevertheless, Casimir energy is a form of Zero Point
Energy (ZPE) which, usually manifests Ultra Violet (UV) divergences. To keep
under control the UV divergences, usually one invokes a standard
regularization/renormalization process. However, an alternative procedure
can be taken under consideration by distorting spacetime since the
beginning. This distortion is better known as Gravity's Rainbow. Since
Gravity's Rainbow switches on at the Planck scale it is likely that ZPE can
be used as a tool to produce a topology change. Note that in Ref.\cite{DGL},
the ZPE was used as an indicator for a topology change without a Gravity's
Rainbow scheme. In this contribution we will explicitly show how Gravity's
Rainbow comes into play to produce a topology change as a ZPE consequence.

\section{Self-sustained Traversable Wormholes}

\label{sec:2}In this Section we shall consider the formalism outlined in
detail in Refs. \cite{Remo,Remo1}, where the graviton one loop contribution
to a classical energy in a wormhole background is used. The spacetime metric
representing a spherically symmetric and static wormhole is given by 
\begin{equation}
ds^{2}=-e^{2\Phi (r)}\,dt^{2}+\frac{dr^{2}}{1-b(r)/r}+r^{2}\,(d\theta
^{2}+\sin ^{2}{\theta }\,d\phi ^{2})\,,  \label{metricwormhole}
\end{equation}%
where $\Phi (r)$ and $b(r)$ are arbitrary functions of the radial
coordinate, $r$, denoted as the redshift function, and the shape function,
respectively \cite{MT}. The radial coordinate has a range that increases
from a minimum value at $r_{0}$, corresponding to the wormhole throat, to
infinity. A fundamental property of a wormhole is that a flaring out
condition of the throat, given by $(b-b^{\prime }r)/b^{2}>0$, is imposed 
\cite{MT,Visser}, and at the throat $b(r_{0})=r=r_{0}$, the condition $%
b^{\prime }(r_{0})<1$ is imposed to have wormhole solutions. Another
condition that needs to be satisfied is $1-b(r)/r>0$. For the wormhole to be
traversable, one must demand that there are no horizons present, which are
identified as the surfaces with $e^{2\Phi }\rightarrow 0$, so that $\Phi (r)$
must be finite everywhere. The classical energy is given by 
\begin{equation*}
H_{\Sigma }^{(0)}=\int_{\Sigma }\,d^{3}x\,\mathcal{H}^{(0)}=-\frac{1}{16\pi G%
}\int_{\Sigma }\,d^{3}x\,\sqrt{g}\,R\,,
\end{equation*}%
where the background field super-hamiltonian, $\mathcal{H}^{(0)}$, is
integrated on a constant time hypersurface. $R$ is the curvature scalar, and
using metric $\left( \ref{metricwormhole}\right) $, is given by 
\begin{equation*}
R=-2\left( 1-\frac{b}{r}\right) \left[ \Phi ^{\prime \prime }+(\Phi ^{\prime
})^{2}-\frac{b^{\prime }}{r(r-b)}-\frac{b^{\prime }r+3b-4r}{2r(r-b)}\,\Phi
^{\prime }\right] \,.
\end{equation*}%
We shall henceforth consider a constant redshift function, $\Phi ^{\prime
}(r)=0$, which provides interestingly enough results, so that the curvature
scalar reduces to $R=2b^{\prime }/r^{2}$. Thus, the classical energy reduces
to 
\begin{equation}
H_{\Sigma }^{(0)}=-\frac{1}{2G}\int_{r_{0}}^{\infty }\,\frac{dr\,r^{2}}{%
\sqrt{1-b(r)/r}}\,\frac{b^{\prime }(r)}{r^{2}}\,.  \label{classical}
\end{equation}%
A traversable wormhole is said to be \textquotedblleft \textit{self sustained%
}\textquotedblright\ if%
\begin{equation}
H_{\Sigma }^{(0)}=-E^{TT},  \label{SS}
\end{equation}%
where $E^{TT}$ is the total regularized graviton one loop energy. Basically
this is given by 
\begin{equation}
E^{TT}=-\frac{1}{2}\sum_{\tau }\left[ \sqrt{E_{1}^{2}\left( \tau \right) }+%
\sqrt{E_{2}^{2}\left( \tau \right) }\right] \,,  \label{OneL}
\end{equation}%
where $\tau $ denotes a complete set of indices and $E_{i}^{2}\left( \tau
\right) >0$, $i=1,2$ are the eigenvalues of the modified Lichnerowicz
operator%
\begin{equation}
\left( \hat{\bigtriangleup}_{L\!}^{m}\!{}\;h^{\bot }\right) _{ij}=\left(
\bigtriangleup _{L\!}\!{}\;h^{\bot }\right)
_{ij}-4R{}_{i}^{k}\!{}\;h_{kj}^{\bot }+\text{ }^{3}R{}\!{}\;h_{ij}^{\bot }\,,
\end{equation}%
acting on traceless-transverse tensors of the perturbation and where $%
\bigtriangleup _{L}$is the Lichnerowicz operator defined by%
\begin{equation}
\left( \bigtriangleup _{L}\;h\right) _{ij}=\bigtriangleup
h_{ij}-2R_{ikjl}\,h^{kl}+R_{ik}\,h_{j}^{k}+R_{jk}\,h_{i}^{k},
\end{equation}%
with $\bigtriangleup =-\nabla ^{a}\nabla _{a}$. For the background $\left( %
\ref{metricwormhole}\right) $, one can define two r-dependent radial wave
numbers%
\begin{equation}
k_{i}^{2}\left( r,l,\omega _{i,nl}\right) =\omega _{i,nl}^{2}-\frac{l\left(
l+1\right) }{r^{2}}-m_{i}^{2}\left( r\right) \quad i=1,2\quad ,  \label{kTT}
\end{equation}%
where%
\begin{equation}
\left\{ 
\begin{array}{c}
m_{1}^{2}\left( r\right) =\frac{6}{r^{2}}\left( 1-\frac{b\left( r\right) }{r}%
\right) +\frac{3}{2r^{2}}b^{\prime }\left( r\right) -\frac{3}{2r^{3}}b\left(
r\right) \\ 
\\ 
m_{2}^{2}\left( r\right) =\frac{6}{r^{2}}\left( 1-\frac{b\left( r\right) }{r}%
\right) +\frac{1}{2r^{2}}b^{\prime }\left( r\right) +\frac{3}{2r^{3}}b\left(
r\right)%
\end{array}%
\right.  \label{masses}
\end{equation}%
are two r-dependent effective masses $m_{1}^{2}\left( r\right) $ and $%
m_{2}^{2}\left( r\right) $. When we perform the sum over all modes, $E^{TT}$
is usually divergent. In Refs. \cite{Remo,Remo1} a standard
regularization/renormalization scheme has been adopted to handle the
divergences. In this contribution, we will consider the effect of Gravity's
Rainbow on the graviton to one loop. One advantage in using such a scheme is
to avoid the renormalization process and to use only one scale: the Planck
scale.

\section{Gravity's Rainbow at work and Topology Change}

One of the purposes of Eq.$\left( \ref{SS}\right) $ is the possible
discovery of a traversable wormhole with the determination of the shape
function. When Gravity's Rainbow is taken under consideration, spacetime is
endowed with two arbitrary functions $g_{1}\left( E/E_{P}\right) $ and $%
g_{2}\left( E/E_{P}\right) $ having the following properties%
\begin{equation}
\lim_{E/E_{P}\rightarrow 0}g_{1}\left( E/E_{P}\right) =1\qquad \text{and}%
\qquad \lim_{E/E_{P}\rightarrow 0}g_{2}\left( E/E_{P}\right) =1.  \label{lim}
\end{equation}%
$g_{1}\left( E/E_{P}\right) $ and $g_{2}\left( E/E_{P}\right) $ appear into
the solutions of the modified Einstein's Field Equations\cite{MagSmo}%
\begin{equation}
G_{\mu \nu }\left( E/E_{P}\right) =8\pi G\left( E/E_{P}\right) T_{\mu \nu
}\left( E/E_{P}\right) +g_{\mu \nu }\Lambda \left( E/E_{P}\right) ,
\label{Gmn}
\end{equation}%
where $G\left( E/E_{P}\right) $ is an energy dependent Newton's constant,
defined so that $G\left( 0\right) $ is the low-energy Newton's constant and $%
\Lambda \left( E/E_{P}\right) $ is an energy dependent cosmological
constant. Usually $E$ is the energy associated to the particles deforming
the spacetime geometry. Since the scale of deformation involved is the
Planck scale, it is likely that spacetime itself fluctuates in such a way to
produce a ZPE. However the deformed Einstein's gravity has only one particle
available: the graviton. As shown in Ref.\cite{RGFSNL}, the self sustained
equation $\left( \ref{SS}\right) $ becomes 
\begin{equation}
\frac{b^{\prime }(r)}{2Gg_{2}\left( E/E_{P}\right) r^{2}}=\frac{2}{3\pi ^{2}}%
\left( I_{1}+I_{2}\right) \,.  \label{ETT}
\end{equation}%
Eq.$\left( \ref{ETT}\right) $ is finite for appropriate choices of the
Rainbow's functions $g_{1}\left( E/E_{P}\right) $ and $g_{2}\left(
E/E_{P}\right) $. We assume that%
\begin{equation}
g_{1}\left( E/E_{P}\right) =\exp \left( -\alpha E^{2}/E_{P}^{2}\right)
\qquad g_{2}\left( E/E_{P}\right) =1\,,  \label{GRF}
\end{equation}%
where $\alpha $ $\in 
\mathbb{R}
$ and $g_{2}\left( E/E_{P}\right) =1$, to avoid Planckian distortions in the
classical term. We find%
\begin{equation}
I_{1}=3\int_{\sqrt{m_{1}^{2}\left( r\right) }}^{\infty }\exp (-\alpha \frac{%
E^{2}}{E_{P}^{2}})E^{2}\sqrt{E^{2}-m_{1}^{2}\left( r\right) }dE\,,
\label{I11}
\end{equation}%
and%
\begin{equation}
I_{2}=3\int_{\sqrt{m_{2}^{2}\left( r\right) }}^{\infty }\exp (-\alpha \frac{%
E^{2}}{E_{P}^{2}})E^{2}\sqrt{E^{2}-m_{2}^{2}\left( r\right) }dE\,.
\label{I22}
\end{equation}%
Following Ref.\cite{RGFSNL}, after the integration one finds that Eq.$(\ref%
{ETT})$ can be rearranged in the following way%
\begin{equation}
\,\frac{b^{\prime }(r)}{2Gr^{2}}=\frac{E_{P}^{4}}{2\pi ^{2}}\left[ \frac{%
x_{1}^{2}}{\alpha }\exp \left( -\frac{\alpha x_{1}^{2}}{2}\right)
K_{1}\left( \frac{\alpha x_{1}^{2}}{2}\right) +\frac{x_{2}^{2}}{\alpha }\exp
\left( \frac{\alpha x_{2}^{2}}{2}\right) K_{1}\left( \frac{\alpha x_{2}^{2}}{%
2}\right) \right] ,  \label{bp}
\end{equation}%
where $x_{1}=\sqrt{m_{1}^{2}\left( r\right) /E_{P}^{2}}$ , $x_{2}=\sqrt{%
m_{2}^{2}\left( r\right) /E_{P}^{2}}$ and $K_{1}\left( x\right) $ is a
modified Bessel function of order 1. Note that it is extremely difficult to
extract any useful information from this relationship, so that in the
following we consider two regimes, namely the cis-planckian regime, where $%
x_{i}\ll 1$ ($i=1,2$), and the trans-planckian r\'{e}gime, where $x_{i}\gg 1$%
. In Ref.\cite{RGFSNL}, it has been shown that the cis-planckian regime does
not produce solutions compatible with traversability. On the other hand when
we fix our attention on the trans-planckian regime, i.e., $x_{1}\gg 1$ and $%
x_{2}\gg 1$, we obtain the following approximation 
\begin{equation}
\frac{1}{2G}\,\frac{b^{\prime }(r)}{r^{2}}\simeq \frac{E_{P}^{4}}{8\sqrt{%
\alpha ^{3}\pi ^{3}}}\left[ \exp \left( -\alpha x_{1}^{2}\right)
x_{1}+O\left( \frac{1}{x_{1}}\right) +\exp \left( -\alpha x_{2}^{2}\right)
x_{2}+O\left( \frac{1}{x_{2}}\right) \right] \,.  \label{Largex}
\end{equation}%
Note that in this regime, the asymptotic expansion is dominated by the
Gaussian exponential so that the quantum correction vanishes. Thus, the only
solution is $b^{\prime }(r)=0$ and consequently we have a constant shape
function, namely, $b(r)=r_{t}$. It is interesting to observe that Eq.$\left( %
\ref{ETT}\right) $ can be interpreted also in a different way. Indeed, if we
fix the background on the r.h.s. of Eq.$\left( \ref{ETT}\right) $ and
consequently let the quantum fluctuations evolve, one can verify what kind
of solutions it is possible to extract from the l.h.s. in a recursive way.
In this way, if we discover that the l.h.s. has solutions which
topologically differ from the fixed background of the r.h.s., we can
conclude that a topology change has been induced from quantum fluctuations
of the graviton for any spherically symmetric background on the r.h.s of Eq. 
$\left( \ref{ETT}\right) $. Of course, this is not a trivial task, therefore
the simplest way to see if a topology change is realized, we fix the
Minkowski background on the r.h.s. of Eq.$\left( \ref{ETT}\right) $. This
means that $b(r)=0$ $\forall r$ and the effective masses become\ 
$ m_{1}^{2}\left( r\right) =m_{2}^{2}\left( r\right) =6/r^{2}.$
Then Eq.$\left( \ref{bp}\right)$ reduces to
\begin{equation}
\frac{1}{2G}\,\frac{b^{\prime }(r)}{r^{2}}=\frac{E_{P}^{4}}{\pi ^{2}}\left[ 
\frac{6}{\alpha \left( rE_{P}\right) ^{2}}\exp \left( -\frac{3\alpha }{%
\left( rE_{P}\right) ^{2}}\right) K_{1}\left( \frac{3\alpha }{\left(
rE_{P}\right) ^{2}}\right) \right] .
\end{equation}%
Let us fix our attention on the trans-planckian regime, i.e., $rE_{P}\ll 1$,
where we can write%
\begin{equation}
\,\frac{b^{\prime }(r)}{r^{2}}\simeq \frac{E_{P}^{2}}{2\sqrt{\alpha ^{3}\pi
^{3}}}\left[ \exp \left( -\alpha \frac{6}{\left( rE_{P}\right) ^{2}}\right) 
\frac{\sqrt{6}}{rE_{P}}+O\left( rE_{P}\right) \right] \,.  \label{TC}
\end{equation}%
Since this is a particular case of Eq.$\left( \ref{Largex}\right) $, we
conclude that the only solution is $b^{\prime }(r)=0$ and consequently we
have a constant shape function, namely, $b(r)=r_{t}$. A comment to the
result $\left( \ref{TC}\right) $ is in order. One could think that Eq.$%
\left( \ref{TC}\right) $ is only a special case of Eq.$\left( \ref{Largex}%
\right) $. Of course this is not true, because Eq.$\left( \ref{Largex}%
\right) $ uses a different initial condition with respect to Eq.$\left( \ref%
{TC}\right) $. Indeed, in Eq.$\left( \ref{Largex}\right) $ the background is
arbitrary, while in Eq.$\left( \ref{TC}\right) $ one considers a Minkowski
line element and the solution is obtained with an iterative process. One can
observe that this procedure could be approached also distorting the one loop
graviton by means of a Noncommutative geometry like in Ref.\cite{RGPN,RemoTM}%
, where the classical Liouville measure is modified into\cite{RGPN}%
\begin{equation}
dn_{i}=\frac{d^{3}\vec{x}d^{3}\vec{k}}{\left( 2\pi \right) ^{3}}\exp \left( -%
\frac{\theta }{4}\left( \omega _{i,nl}^{2}-m_{i}^{2}\left( r\right) \right)
\right) ,\quad i=1,2.  \label{moddn}
\end{equation}%
$m_{i}^{2}\left( r\right) $ are the effective masses described in $\left( %
\ref{masses}\right) $ and $\theta $ is the Noncommutative parameter. While
nothing can be said about the effect of Noncommutative geometry on topology
change, a result can be extracted from the traversability of the wormhole.
Indeed, if one fixes the form of the shape function to be $b\left( r\right)
=r_{0}^{2}/r,$which is the prototype of the traversable wormholes\cite{MT},
one gets $r_{0}=0.28l_{P},$with $\theta $ fixed at $\theta =7.\,\allowbreak
43\times 10^{-2}l_{P}^{2}$. If we compare the result obtained in Ref.\cite%
{RGFSNL} using Gravity's Rainbow, one finds the following value for the
radius $r_{t}=1.\,\allowbreak 46l_{P}$,$\,$which is slightly larger than $%
r_{0}$. The conclusion is that Gravity's Rainbow and Noncommutative geometry
keep under control the UV divergences in this ZPE calculation connected with
the self-sustained equation. In both cases we find that the result is a
Wheeler wormhole. This means that, from the point of view of traversability,
the wormhole will be traversable in principle, but not in practice.

\end{document}